\documentclass{ar2e}

\usepackage[numbers,sort&compress]{natbib}
\usepackage{amsmath}
\usepackage{graphicx}

\begin{document}

\jname{Annu. Rev. Cond. Matt. Phys.}
\jyear{2014}
\jvol{??}
\ARinfo{???}

\title{The Statistical Physics of Athermal Materials}

\markboth{Bi, Henkes, Daniels, Chakraborty}{The Statistical Physics of Athermal Materials}

\author{
Dapeng Bi \affiliation{Department of Physics, Syracuse University \\ Syracuse, 13244, USA}
Silke Henkes \affiliation{ICSMB, Department of Physics, University of Aberdeen \\ Aberdeen, AB24 3UE, UK}
Karen E. Daniels \affiliation{Department of Physics, North Carolina State University \\ Raleigh, NC 27695, USA}
Bulbul Chakraborty \affiliation{Martin Fisher School of Physics, Brandeis University \\ Waltham, MA 02454, USA}
}

\begin{keywords}
granular materials, statistical physics, athermal
\end{keywords}

\begin{abstract}
At the core of equilibrium statistical mechanics lies the notion of statistical ensembles: a collection of microstates, each occurring with a given {\it a priori} probability that depends only on a few macroscopic parameters such as temperature, pressure, volume, and energy.  In this review article, we discuss recent advances in establishing statistical ensembles for athermal materials.  The broad class of granular and particulate materials is immune from the effects of thermal fluctuations because the constituents are macroscopic.  In addition, interactions between grains are frictional and dissipative, which invalidates the fundamental postulates of equilibrium statistical mechanics.  However,  granular materials exhibit distributions of microscopic quantities that are reproducible and often depend on only  a few macroscopic parameters.  We explore the history of statistical ensemble ideas in the context of granular materials,  clarify the nature of such ensembles and their foundational 
principles,  highlight advances in testing key ideas, and discuss applications of ensembles to analyze the collective behavior of granular materials.
\end{abstract}

\maketitle


\section{Introduction \label{s:intro}}

Athermal, cohesionless particulate materials such as dry grains and dense, non-Brownian suspensions exhibit  features that are generic to many non-ergodic, disordered (glassy) 
systems~\citep{Bouchaud_lecture, Bouchbinder:2007qf}. 
However, these granular materials  are non-ergodic in the extreme sense, as they stay in a single configuration unless driven externally.   Exploration of configuration space is completely controlled by the driving protocol, and since  cohesive forces are absent,  fluid or solid-like  behavior emerges purely as a response  to external driving~\citep{Cates1999Jamming-and-str,Cates1998}.  These systems are, thus, the epitome of athermal condensed matter.  The objective of this article is to describe recent developments in constructing statistical ensembles for such athermal systems.  We examine the history of statistical ensembles for granular materials, which began with the ideas of Sam Edwards~\citep{Edwards1989}, describe recent developments in theory and 
experiments, and explore future applications.

Although thermal fluctuations are absent, granular systems nonetheless exhibit large fluctuations~\citep{Springer}.  This is exemplified by the exponential distribution of contact forces in jammed (solid-like) granular assemblies~\citep{Liu1995,majmudar05a}, and large,  intermittent stress fluctuations in quasistatic flows~\citep{Howell1999}.  The ubiquitous presence of fluctuations with well-defined {\it distributions} that seem to depend on a handful of macroscopic parameters~\citep{Springer} suggests that statistical ensembles could prove to be an important tool for predicting the
emergent properties of granular materials.  There are two aspects that need to be elucidated; (i) the establishment of the ensemble and (ii) how to use the ensembles to calculate and predict the emergent behavior of granular materials. After introducing the history of athermal ensembles in \S\ref{s:intro}, we discuss the special role of constraints in the   enforcement of mechanical equilibrium, and how this differs from thermal ensembles. In 
\S\ref{s:tests}, we review the existing experimental and numerical tests of the ensembles and in 
\S\ref{s:statmech} describe how to use the distributions of microstates in an ensemble to calculate collective properties. These same techniques can be extended to systems with slow dynamics, and examples of this are provided in \S\ref{s:dynamics}. Finally, we close with a summary of open questions in the field (\S\ref{s:questions}).

\subsection{History}

Twenty-five years ago, Sam Edwards and coworkers \citep{Edwards1989} considered the problem of what sets the packing density of granular materials under simple operations such as shaking, stirring or compression.  They proposed a statistical ensemble approach, which was formulated along lines paralleling that of equilibrium statistical mechanics and thermodynamics.
In equilibrium statistical mechanics, conservation of energy ($1^\mathrm{st}$ Law) implies that the total energy $E$ of an isolated system  cannot fluctuate.  The set of microstates ($\nu$) with a given total energy $E$, calculated from the positions and momenta of all particles, form  the microcanonical ensemble.
The macroscopic, extensive variable $E$ characterizes the ensemble, and all microstates with this energy are assumed to occur with equal probability.  Different physical systems are distinguished by the density of states $\Omega_B(E) = \sum_{\nu} \delta (E-E_{\nu})$.  
Isolated systems are important only for establishing the foundations of thermodynamics, and the {\it canonical} ensemble describing a system in contact with a heat bath is characterized by an intensive macrosopic variable: the temperature $\frac{1}{T} = \frac{\partial S(E)}{\partial E}$.  It is useful to summarize the postulates used to arrive at the framework described above.  In addition to energy conservation, which is dynamical in origin, the fundamental assumption underlying statistical mechanics is the equiprobability of microstates with a given energy, which also follows from the maximization of entropy \citep{chandler_book}. 

Edwards and collaborators set about formulating an analogous theory of granular materials.
Due of frictional forces and dissipation, energy is not conserved in granular systems, it is therefore not an appropriate state variable.
There are two aspects to the Edwards formulation.  One that is often forgotten is the assertion that the dynamics of slowly driven  granular systems is controlled by the statistical properties of blocked states~\citep{Edwards2002}. For infinitely rigid grains, these blocked  microstates are composed of grains in mechanical equilibrium such that grains cannot be moved without causing a finite overlap.  In the modern parlance, these would be called jammed states \citep{Donev2004Jamming-in-hard,Liu-2010-JTM}.  Motivated by the central role that the packing density plays in determining the nature of jammed states, Edwards proposed to replace the Hamiltonian or energy function by a volume function and defined the analogous density of states: $\Omega(V)=\sum_{\nu} \delta(V-V_{\nu})$, where $\nu$ is now restricted to {\it jammed} states, and is defined by just the positions of grains.   Following the steps of the 
development of thermodynamics, the Edwards entropy is then defined as 
$S \equiv \lambda \ln \Omega(V)$, which yields a canonical ensemble characterized by a corresponding temperature-like variable $X$:
\begin{equation}
\frac{1}{X}  = \frac{\partial S(V)}{\partial V},
\end{equation}
which Edwards named {\itshape compactivity}. Just as the Boltzmann constant $k_B$ gives the ordinary entropy the correct units of J/K, the constant $\lambda$ sets the units of compactivity relative to volume. Similarly, one can write quantities analogous to the specific heat, the Boltzmann distribution, etc. A summary of these ordinary statistical mechanical quantities, and their translations into the Edwards ensemble, is given in Table.~\ref{t:translate}.   Although the Edwards ensemble looks exactly like the ensembles of equilibrium statistical mechanics, it is different in one very important way: {\it it is constrained to consider only those states that are blocked or jammed}.  Any calculation involving sums over states such as a partition function has to explicitly impose this constraint.   

\begin{table}[b]
 \begin{center}
\begin{tabular}{|c|c|c|c|}
\hline
~ 
     & {\bf Boltzmann} 
     & {\bf Edwards} 
     & {\bf generalized Edwards} \\ \hline
conserved quantity 
     & energy, $E$ 
     & volume, $V$ 
     & force moment tensor, $\hat \Sigma$ \\
\# of valid configurations 
     & $\Omega_B(E)$
     & $\Omega(V)$ 
     & $\Omega(\hat \Sigma)$ \\
entropy 
     & $S = k_B \ln \Omega_B$
     & $S = \lambda \ln \Omega$  
     & $S = \lambda \ln \Omega$ \\ 
equilibrating quantity 
     & temperature
     & compactivity
     & angoricity tensor \\
~ 
     & $\frac{1}{T} = \frac{\partial S(E)}{\partial E}$ 
     & $\frac{1}{X} = \frac{\partial S(V)}{\partial V}$
     & $\alpha_{\mu \lambda} = \frac{\partial S({\Sigma_{\mu \lambda}}) }{\partial {\Sigma_{\mu \lambda}}}$\\
distribution 
     & $ ^{-\frac{E}{k_B T}}$ 
     & $e^{-\frac{V}{X}}$ 
     & $e^{-{\hat \alpha} : {\hat \Sigma}}$  \\ \hline
 \end{tabular}
 \end{center}
\caption{Table of similar quantities from ordinary statistical mechanics and the original and generalized Edwards ensemble. In practice, $\lambda$ is often to be taken to have a value unity with units of 1/Volume (1/Stress for the generalized ensemble), for simplicity. Note that the angoricity, unlike the other temperatures, is defined without a reciprocal. \label{t:translate}}
\end{table}

The name compactivity was chosen because it describes how far a system is from its densest possible state, beyond which no further compaction is possible. The extremal values of $X$ are chosen to give a sensible physical interpretation of the limits of validity of the ensemble. The lowest possible compactivity ($X=0$) has come to be identified with random close packing, the densest packing for which no crystalline order is present~\citep{Bernal1960, Song-2008-PDJ, ciamarra_statistical_2012}.  The upper limit, at $X=\infty$, is typically identified  with random loose packing \citep{Onoda-1990-RLP, Jerkins2008}, the loosest packing for which mechanical stability is still present. Beyond these values, the ensemble is no longer defined.

\begin{figure}
\centerline{\includegraphics[height=2in]{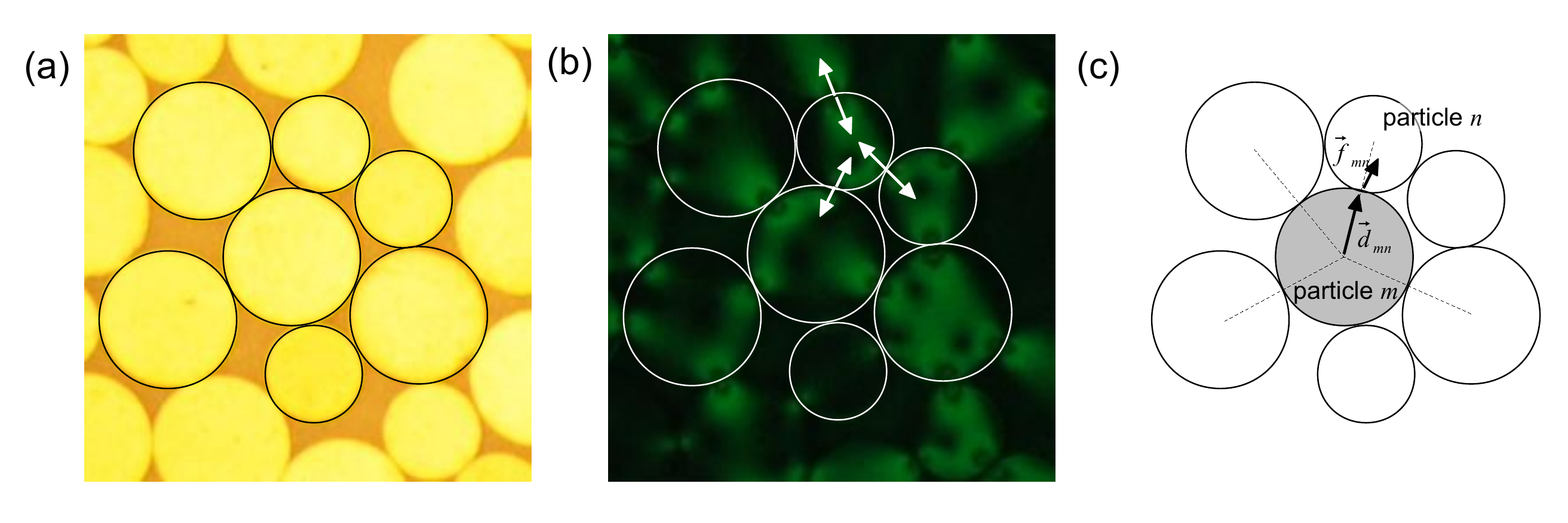}}
\caption{Image from experiments on two-dimensional photoelastic disks. (a) Particles in unpolarized (b) polarized light. Adapted from \citep{Puckett2013}. From (b), it is possible to extract vector contact force ${\vec f}_{mn} = -{\vec f}_{nm}$ at each interparticle contact, shown schematically as the white arrows. (c) Schematic of calculation of force moment tensor $\hat \Sigma$ according to Eq.~\ref{fmt_def}, using $f_{mn}$ and the vector ${\vec d}_{mn}$ pointing from the center of particle $m$ to the interparticle contact with particle $n$. 
Dashed lines connect the center of particle $m$ with the center of each particle it is in contact with. 
\label{f:exp-forces}}
\end{figure}

Soon after the introduction of the Edwards ensemble, it was realized that contact forces needed to be incorporated into the definition of jammed microstates. Each grain must locally satisfy force and torque balance, and in the presence of friction these forces cannot be deduced just from positions of grains.   There was also a heightened awareness \citep{Liu1995,Howell1999} of the significance of earlier observations \citep{Dantu1957, Drescher-1972-PVM} of what have come to be known as {\itshape force chains}. These are roughly co-linear chains of particles through which larger-than-average stresses are transmitted. These considerations led to the proposal of a generalized ensemble \citep{Edwards2005, Blumenfeld-2009-GSS, Henkes2005} with stress as an additional governing state variable; the corresponding temperature-like variable was named {\itshape angoricity} by Edwards, from the 
modern Greek word $\acute{\alpha} \gamma \chi o \varsigma$, meaning ``stress.''  There is a growing consensus that the force moment tensor $\hat \Sigma$ is  the extensive variable most closely analogous to energy in equilibrium statistical mechanics:
\begin{equation}
\hat \Sigma = \sum_{m,n} \vec d_{mn} \vec f_{mn} ~,
\label{fmt_def}
\end{equation}
where ${\vec d}_{mn}$ is the vector pointing from the center of particle $m$ to the interparticle contact with its neighbor particle $n$, ${\vec f}_{mn}$ is the interparticle contact force, and the sum is over all pairs of grains. These quantities are illustrated in Fig.~\ref{f:exp-forces}, along with sample images illustrating how they are measured in experiments.  Unlike the volume $V$, $\hat \Sigma$ depends on both the positions of grains and their interparticle contact forces.
Because the force moment is a tensorial quantity, the angoricity is a tensor as we well: 
\begin{equation}
{\alpha_{\mu \lambda}} \equiv \frac{\partial S({ \hat \Sigma}) }{\partial {\Sigma_{\mu \lambda}}}.
\label{alpha_def}
\end{equation}
As the $\mu, \lambda$ component of the force moment tensor goes to zero (corresponding to a loss of rigidity),  we expect the corresponding angoricity to tend to to infinity. Since the original proposal, a number of different formulations for the stress ensemble have been published \citep{Edwards2005, Blumenfeld-2009-GSS, 
Henkes2009,Henkes2007,Metzger2008,Tighe:2008ye}.

\subsection{Athermal vs. Thermal Ensembles}

The volume and stress ensembles described in the previous section have come to be collectively known as generalized Edwards ensembles \citep{Chakraborty2010}, and are distinct from ordinary statistical mechanical ensembles in  important ways.
First, they are completely non-ergodic. Because there is no temporal evolution, there is no equivalence of time averages and ensemble averages.
This peculiarity arises because granular materials are athermal: at room temperature, the thermal energy is many orders of magnitude lower than required to move one grain over another against the force of gravity. For example, a millimeter-sized grain of sand  would require $mgd \approx 10^{-5}$~J to move upward its own diameter against the force of gravity. At room temperature (300 K), this corresponds to $\approx 10^{15} \, k_B T$ and therefore thermal motion is statistically negligible. As such, all granular systems require external driving to explore new configurations, and the Edwards ensemble (or its variants) can only be interpreted as an ensemble of  microstates created by independent realizations 
with certain macroscopic parameters held fixed. Another type of granular ensemble is that of subsystems within a single thermodynamically large system.  In this context, ``thermalization'' refers to the establishment of equal angoricity and compactivity in these different subsystems.  This granular thermalization is distinct from the establishment of effective temperatures in glassy systems \citep{Cugliandolo1999Thermal-propert,Langer2007Steady-state-ef}, where the effective temperatures encode the fluctuations of the slowly-relaxing modes, and are dynamical in nature.

While the assumption of equiprobability  of all valid microstates is well established for equilibrium statistical mechanics,  the situation is less clear for the Edwards ensembles.   In fact, tests of the hypothesis for jammed states in experiments, simulations, and exactly solvable models reveal that equiprobability is not universally valid, but does hold under some conditions \citep{Barrat2001Edwards-measure,Barrat2000Edwards-measure,Coniglio2002Probability-dis,Coniglio2001Applications-of,Kurchan2001Recent-theories,Makse2002}.  In particular, \citet{Gao2009} used closely-matched simulations and experiments to generate millions of statistically-independent packings of seven disk-shaped particles. By enumerating all observed unique configurations, they observed that the same finite number of microstates was observed in both simulations and experiments, and that the relative frequencies of the these microstates was highly nonuniform (they differ by factors of up to a million). However, equiprobability is {\
itshape not} essential for either the definition of compactivity \citep{McNamara-2009-MGE} or angoricity \citep{Henkes2009}.  Furthermore, it has recently been shown~\citep{Asenjo2014} that the Edwards entropy can satisfy extensivity even without equiprobability, and that a factor of $1/N!$ is necessary to avoid an athermal version of the Gibbs paradox.

If configurations are not equiprobable, one can generalize the definition of $\Omega(V)$ to include the weights of microstates: $\Omega(V) = \sum_{\nu} \omega_{\nu} \, \delta(V - V_{\nu})$.
The much weaker condition of factorability, $\Omega(V_1 + V_2) = \Omega(V_1) \Omega (V_2)$, is necessary and sufficient  for defining angoricity and compactivity 
\citep{Bertin2006Definition-and-,Bertin2005Nonequilibrium-,Bertin2004Temperature-in-}.
An important point to note is that, unlike $\Omega_B (E)$, $\Omega (V)$ is not
determined by any Hamiltonian.  A pragmatic perspective is to assert that
different protocols lead to different  microstate probability $\omega_{\nu}$ and therefore, to different $\Omega (V)$.  The ensemble approach is applicable, but each
protocol would be like choosing a new Hamiltonian that defines the density of states.

\subsection{Conservation Principles \label{s:conservation}}

The objective of a statistical ensemble framework is to predict the probability of occurrence of a microscopic configuration, given a set of macroscopic constraints.   In equilibrium statistical mechanics, there are natural macroscopic constraints that emerge from conservation laws:  energy, volume and number of particles.  In jammed, athermal systems, there are similarly a natural set of macroscopic variables and their associated conservation principles.

Any microscopic jammed state of $N$ grains is completely specified by the positions of the grains and the forces (including frictional forces) at inter-grain contacts: $\nu = \left\lbrace \vec r_m,~  \vec f_{mn} \right\rbrace$, where $\vec r_m$ is the position of the $m$-th grain and $\vec f_{mn}$ is the contact force between grains $m$ and $n$.   As in thermal systems, and in other jammed or glassy systems, a granular packing can exist in many different microscopic states for a fixed set of macroscopic variables such as volume or applied stress.  In addition, there is a microscopic indeterminacy \citep{Bouchaud_lecture} arising from the Coulomb condition for static friction which imposes an inequality constraint: $|f_T| \leq \mu |f_N|$ such that the magnitude of the tangential force  $\vec f_T$ cannot exceed the static friction coefficient $\mu$ times the magnitude of the normal force $\vec f_N$.  Thus, there is a range of tangential forces that are allowed for any  {\it microscopic} grain configuration.  
What, 
then, is a natural 
microcanonical ensemble  for jammed states?  

To address this question, we start by examining the constraints that must be satisfied by jammed (blocked) microstates.  For infinitely rigid grains, there is a strict constraint on the $\lbrace \vec r_m \rbrace$ since grains cannot overlap. The jammed states in the original Edwards ensemble  are then all of the ``just-touching'' configurations, and the volume occupied by $N$ grains is a macrosopic invariant: it is {\itshape globally} conserved.  For grains that are not infinitely rigid, the non-overlapping constraint vanishes, and it is no longer clear what volume is conserved.  Of course, one can fix the total volume $V$ occupied by $N$ grains through the nominal packing fraction, but this is not a consequence of any constraints on jammed states.  The constraints that remain are those of mechanical equilibrium, which involve the contact forces $\left\lbrace \vec f_{mn} \right\rbrace$.
These constraints lead to a robust conservation law involving the force-moment tensor~\citep{Henkes2007,Blumenfeld2008Notes,Ball2002} of the granular packing, $\hat{\Sigma}$, defined in Eq.~\ref{fmt_def},
which is related to the Cauchy stress
tensor~\citep{goddard86} $\hat{\sigma}$:
\begin{equation}
\hat \Sigma = V \hat \sigma 
\label{fmt_def2}
\end{equation}

The conservation of the force-moment tensor arises purely from {\it local} constraints of force and torque balance. It holds for any material that is in mechanical equilibrium, regardless of dimensionality and microscopic details. Here, we will demonstrate this conservation principle in a generic two dimensional $T=0$ solid using a continuum stress field, but a formulation that explicitly incorporates the discreteness of the granular assembly can also be constructed~\citep{DeGiuli2011,Henkes2009}.

We consider a two-dimensional (2D) $T=0$ solid under periodic boundary conditions (PBC). Mechanical equilibrium requires that the continuum stress field $\hat{\sigma}(\bf{r})$ obeys force and torque balance everywhere. 
In the absence of body forces, the continuum stress field satisfies
\begin{equation}
\vec{\nabla} \cdot \hat{\sigma}({\vec{r}})= 0 ~
\label{divergencefree}
\end{equation}
due to force balance. Local torque balance also requires $\sigma_{\mu \lambda}=\sigma_{\lambda \mu}$. There is an additional constraint of $\sigma_{\mu \mu}>0$ if the system of interest contains non-cohesive interactions (e.g. dry grains), for which tensile forces are not possible.

\begin{figure}
\begin{center}
\includegraphics[width=\textwidth]{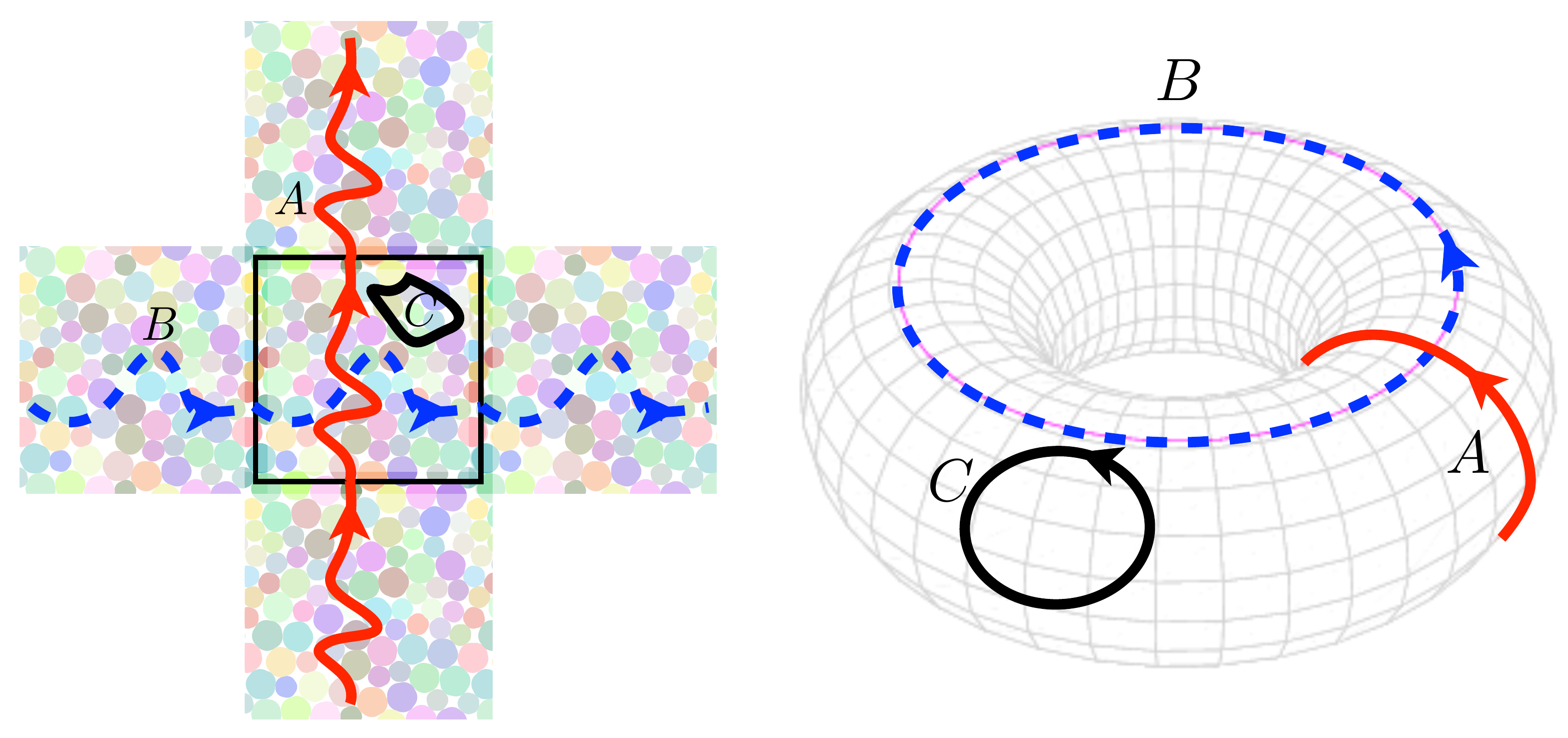}
\caption
{
Illustration of Invariants:  
(a) The physical system, a 2D granular solid of size $L \times L$, is outlined by the black box, also shown are its images under PBC. The lines $A$ \& $B$ represent the two distinct classes of non-contractible loops in the system and $C$ represents a trivial loop. 
(b) Representation of the system on the surface of a torus. Loops $A$ \& $B$ are non-contractible and correspond to the same labeling as in (a). To change $\vec{F}_x$ (or $\vec{F}_y$), a change has to be made on the non-contractible loop $B$ (or $A$).
}
\label{loops}
\end{center}
\end{figure}

For a 2D system under PBC (equivalent to the surface of a torus), there are three classes of topologically distinct loops, 
labeled $A,B~\&~C$ in Fig.~\ref{loops}.  
For any region enclosed by a topologically trivial (contractible) loop (type $C$), the total force acting on it can be calculated by applying the divergence theorem to the stress field 
\begin{equation}
\oint_{C} dS \ \hat{\sigma} \cdot \hat{n}=\int dV \ {\vec \nabla} \cdot \hat{\sigma}= 0,
\label{trivial_loop}
\end{equation}
yielding the same result regardless of loop shape, which simply reflects force balance. 
There are also two classes of non-contractible loops: loops labeled $A$ that wrap around the torus in the $x$-direction and those labeled $B$ that wrap around in the $y$-direction.
Applying the divergence theorem leads to two vector quantities 
\begin{equation}
\begin{split}
		\vec{F}_x &= \oint_{A} dS \ \hat{\sigma} \cdot \hat{n} \\
		\vec{F}_y &= \oint_{B} dS \ \hat{\sigma} \cdot \hat{n}
\end{split}
\label{F_vec_def}
\end{equation}
which do not depend on the particular paths loops $A$ and $B$ take. These vectors are topological invariants in the sense that all loops belonging to the same topological class, $A$ or $B$,  have  the same value of $\vec{F}_x$ or $\vec{F}_y$, respectively. 
For a granular system, which is inherently discrete, the loops are not, in the strict sense, continuously deformable since they have to pass through force-bearing contacts.  
While the deformation of loops must obey the constraints of the underlying contact network, 
this does not change the topological invariance of $(\vec {F}_x, \vec {F}_y)$.

For each dimension (here, there are two: $x,y$), there is an invariant vector $\vec F_i$, which represents 
how much force is `propagated' through the material. 
These $\vec F$-vectors are equivalent to the sum of Cauchy traction vectors on a surface in continuum mechanics \citep{liu2002continuum} and are also a formal generalization of the concept of Cartesian load proposed in Ref. \citep{Metzger2008}. The $\vec F$-vectors are related to the force moment tensor which is defined for discrete systems in Eq.~\ref{fmt_def} and in continuum as $\hat{\Sigma}= \int dV \  \hat{\sigma}({\vec r})$ via
\begin{equation} 
\hat{\Sigma} 
	= 
	\left( 
	\begin{array}{cc} 
		L_x & 0 \\ 
		0 & L_y
  	\end{array} 
	\right) 
   	\left(
   	\begin{array}{cc} 
	\vec{F_x}  \cdot \hat{x} &~ \vec{F_x} \cdot \hat{y} \\ 
	\vec{F_y}  \cdot \hat{x} &~ \vec{F_y} \cdot \hat{y} 
        \end{array} 
  	\right) \, .
\label{f_sigma_relation}
\end{equation}
The extensive quantity $\hat \Sigma$ has been used to construct the stress ensemble framework.  However, from the preceding discussion, the natural ``conserved'' variables appear to be $(\vec {F}_x, \vec {F}_y)$ \citep{Chakraborty2010}, which scale linearly with system size.  This feature is reminiscent of classical lattice models, which have  ground state degeneracy that leads to finite entropy.  In these models, the subextensivity of the conserved quantity has nontrivial implications for phase transitions \citep{Dasgupta}.  It would be interesting to explore a theory of grains based on  the conservation of $(\vec {F}_x, \vec {F}_y)$.

A natural microcanonical ensemble for jammed states of grains is one with a fixed value of the force moment tensor, which plays the role of energy in equilibrium statistical mechanics.   The force-moment tensor, $\hat \Sigma$,  is an extensive quantity (i.e. one that scales with
system size), and one can define the generalized density of states: $$
\Omega(\hat \Sigma) = \sum_{\nu} \omega_{\nu} \,  \delta(\hat \Sigma_{\nu} -\hat \Sigma) ~.
$$ The constraints on the microstates $\nu$ are that (i) forces and torques are balanced on every grain, (ii) the Coulomb condition for static friction is satisfied, and (ii) all forces are positive.   If one now assumes factorization: $\Omega(\hat \Sigma_1 + \hat \Sigma_2) = \Omega(\hat \Sigma_1) \Omega (\hat \Sigma_2)$, there is an exact analog of temperature, which is the tensorial angoricity  $\hat \alpha$:
\begin{equation}
\hat \alpha (\hat \Sigma)= \frac{\partial \ln \Omega (\hat \Sigma)}{\partial \hat \Sigma} 
\end{equation}  
as was specified by Eq.~\ref{alpha_def}. 

The preceding derivation is easily generalized to three dimensions. Contractible loops become closed volumes and the three types of non-contractible surfaces correspond to the three invariants  $(\vec {F}_x, \vec {F}_y, \vec {F}_z)$; see also section \S\ref{3D}.

\subsection{Microcanonical and Canonical Formulations}

The original Edwards ensemble can  be generalized to one where $V$ and $\hat \Sigma$ define a microcanonical ensemble, but with no assumption of equiprobability.
This generalized Edwards ensemble has much in common with statistical frameworks developed for glassy systems~\citep{Bouchaud_lecture}.  Collective properties in both are determined by the complexity of the landscape of metastable states.  For example, complexity in spin glasses is measured by the logarithm of the number of local free energy minima that have a given free energy density~\citep{Bouchaud_lecture,Mezard1999}.  An analogous definition applies to the potential energy landscape of supercooled liquids~\citep{Stillinger,Sastry2001}.  For granular systems,  complexity is measured by the generalized entropy, $\Omega(V, \hat \Sigma)$ with the caveat that the  microstates in the Edwards ensemble have to satisfy the constraints of mechanical equilibrium at zero temperature.

The microcanonical formulation of the generalized Edwards ensemble, which does not assume equiprobability, still assumes that the entropy is extensive (factorization of $\Omega$).   Recent numerical simulations establish extensivity in the absence of equiprobability \citep{Asenjo2014}.   What the microcanonical formulation implies is the following:  in a large system of grains that are jammed in a global volume $V_g$ with a global  force-moment tensor $\hat \Sigma_g$, the probability of finding a microstate within a subsystem is given by the Boltzmann-like distribution,  
\begin{equation}
P(\nu = \lbrace {\vec r_m} ~, {\vec f}_{mn} \rbrace) = \frac{1}{Z} \, 
e^{ -\frac{V(\lbrace {\vec r}_m \rbrace)}{X}} \,
e^{ - \hat \alpha: \hat \Sigma(\lbrace {\vec r}_m ~, {\vec f}_{mn}\rbrace)} ~,
\label{can_disbn}
\end{equation}
where $X$ and $\hat \alpha$ depend only on $V_g$ and $\hat \Sigma_g$.  This is a remarkably strong constraint on the probability of microstates in a non-equilibrium system.   In general, this probability can depend on the complete history of preparation (the protocol) and any number of mesoscopic and macroscopic variables.   Rigorous tests of Eq. \ref{can_disbn} are needed to establish the framework for both the microcanonical and canonical ensembles. In the next section, we discuss  such tests for both experimental and numerical systems. 

Defining the ensemble for static grains is the conceptually difficult part.  Calculating the partition function is technically difficult because of the constraints that one still needs to take into account.   An important point to make here is that establishment of a Boltzmann-like distribution does not  imply that the distribution of individual grain volumes or individual contact forces is  exponential \citep{Liu1995}.  This would be true only if granular materials were like ideal gases with no correlations.   As we will discuss in section \S\ref{s:statmech}, granular materials are actually ``strongly interacting'' systems where the interactions arise from constraints rather than interaction potentials.   


\section{Testing the Ensembles \label{s:tests}}

In spite of thought-experiments proposed by theorists \citep{Edwards2002, Bouchaud_lecture}, experimentalists were slow to warm to performing laboratory tests of the applicability and validity of the ideas put forth by Edwards. The earliest approach which gained traction was to consider analogs of the specific heat \citep{Nowak-1998-DFV}. A well-known result of ordinary statistical mechanics is that the energy fluctuations of a system around its mean value $\bar E$ are related in a simple way to its specific heat at constant volume, via the relation $c_V = \left( \frac{\partial {\bar E}}{\partial T} \right)_V = \frac{\langle (E - {\bar E})^2 \rangle}{k_B T^2}$. Similarly, there is an analogous quantity 
$c = \left( \frac{\partial {\bar V}}{ \partial X} \right) = \frac{\langle (V - {\bar V})^2 \rangle}{X^2}$ for the Edwards ensemble which allows you to measure relative changes in compactivity via observations of the volume fluctuations:
\begin{equation}
\int_{V_1}^{V_2} \frac{d {\bar V}}{\langle (V - {\bar V})^2 \rangle} 
    =  \int_{X_1}^{X_2} \frac{d X}{X^2}  =  \frac{1}{X_1} - \frac{1}{X_2} .
 \label{e:specificheat}
\end{equation}
Here, the Boltzmann-like $\lambda$ factor has been dropped for simplicity. This relation was first used  to measure the compactivity of a granular material subject to tapping \citep{Nowak-1998-DFV} and later on static packings prepared through varying degrees of fluidization \citep{Schroter-2005-SSV}. A disadvantage of this volume {\itshape fluctuation method} is that it integrates from a state with known $1/X_1$; typically this is taken to be random loose packing, where $X_\mathrm{RLP} = \infty$. However, it does provide a way to measure the compactivity using only bulk measurements.

As a byproduct, the measurements of the volume fluctuations themselves provide a probe of the Edwards ensemble in the sense that a larger number of configurations corresponds to a larger  ${\langle (V - {\bar V})^2 \rangle}$. This susceptibility can be informative in its own right, as a hallmark of the approach to a critical point. An observation of a minimum in the fluctuations as a function of packing density 
\citep{Schroter-2005-SSV} has since been 
associated with the dilatancy transition \citep{Schroter-2007-PTS,Metayer2011}. Changes in such fluctuations have been observed in a variety of both experiments and simulations of static packings \citep{Aste-2008-STG, Briscoe:2008fk, Schroder-Turk2010, PicaCiamarra2006, Pugnaloni2010}, but also in some systems with  dynamics \citep{Daniels-2006-CFM, Puckett-2011-LOV}. This last case is a test of Edwards' original idea that ensembles could describe  driven granular systems, and therefore hints at the validity of this aspect of the Edwards ensemble beyond strictly jammed states.

A second technique for measuring compactivity  was proposed by \citet{Dean2003}. They observed that while the density of states $\Omega(V)$ and the partition function $Z(X)$ may be unknown for a given granular ensemble, $\Omega(V)$ is identical for the same system under two different conditions. For any particular macroscopic volume $V$, the probability $P$ of observing that volume is given by
\begin{equation}
P(V) = \frac{\Omega(V)}{ Z(X) } e^{-\frac{V}{X}}.
\label{e:overlapping1}
\end{equation}
If you have measurements of $P(V)$ for the same system under two different preparations (enumerated 1, 2), the ratio of those two histograms will be 
\begin{equation}
\frac{P_1 (V)}{P_2 (V)} = 
\frac{Z(X_2)}{Z(X_1)} e^{(\frac{1}{X_2} - \frac{1}{X_1})V}.
\label{e:overlapping2}
\end{equation}
Therefore, by taking logarithm of the ratio and fitting the data to determine the slope $\frac{1}{X_2} - \frac{1}{X_1}$, it is possible to measure changes in the compactivity. If you have a reference point (again, commonly $X_\mathrm{RLP} = \infty$), then this can also be used as an absolute measurement. This technique was first realized by \citet{McNamara-2009-MGE}, who measured these quantities in both simulations and experiments using local Vorono{\"i} volumes calculated from tomographs of an assembly of millimetric glass spheres. Because the histograms were nearly Gaussian, it was difficult to make quantitative conclusions beyond a verification that the technique worked: values of $X$ measured using either Eq.~\ref{e:specificheat} or \ref{e:overlapping2} were in approximate agreement. This same {\it overlapping histogram method} was used on disk-shaped particles in a quasi-two-dimensional packing \citep{Zhao2012}, where it was observed to be properly intensive for assemblies of at least 200 particles \
citep{Zhao2014}. 
Note that this technique for measuring compactivity does not  rely on equiprobability, and is insensitive to it  since the factor $\Omega(V)$ in Eq.~\ref{e:overlapping2} cancels out whatever its form.

Three other methods of calculating $X$ are available for evaluation.
First, using the histograms of local (Vorono{\"i}) volumes, it is possible to measure $X$ directly by fitting a Gamma distribution and considering the ratio of the mean free volume to the shape factor of the Gamma distribution \citep{Aste-2008-EGD}. 
Second, an equation of state relating  the structural degrees  of freedom and the number of independent boundary forces provides a fourth means of measuring $X$, for isostatic packings  \citep{Blumenfeld2003Granular-entrop,Blumenfeld2012}. 
A recent comparison of these two methods shows that they are not necessarily in quantitative agreement with the techniques described by Eq.~\ref{e:specificheat} and \ref{e:overlapping2} \citep{Zhao2014}. 
Finally, it is possible, through simulations, to both count particle configurations and to take the limit $T \rightarrow 0$ in a thermal system. Using such methods on a simulation of slowly-sheared particles, the thermal $T_\mathrm{eff}$ (measured via the diffusivity and the fluctuation-dissipation theorem) was found to compare favorably with the compactivity \citep{Makse2002}.

While these experiments have demonstrated the utility of the Edwards formulation as a description of a system, this was merely a confirmation that compactivity was measurable, not that it represented a truly temperature-like quantity. A more stringent test, as for ordinary thermodynamics, lies with the 0$^\mathrm{th}$ Law of thermodynamics. A minimal expectation of a temperature-like quantity is that it have the ability to equilibrate between a subsystem and a bath. In experiments on a low-friction subsystem within a high-friction bath, the desired result would be for the compactivity to take on the same value in both. However, for a system of two-dimensional disks floated on a frictionless surface and biaxially compacted, it was observed that while $X$ measured using either  Eq.~\ref{e:specificheat} or \ref{e:overlapping2} matched either for the subsystem or for the bath, the two values systematically failed to agree between the subsystem and the bath \citep{Puckett2013}. This represents a significant 
failure of the volume 
ensemble and compactivity. 

Tests of the stress ensemble have a shorter history, but the angoricity appears to be more promising as a valid temperature-like quantity. Because it is a tensor (see Eq.~\ref{alpha_def}), it is helpful to calculate it by considering shear ($\Sigma_\tau$) and compressional ($\Sigma_p$) components separately. Each of these components gives rise to its own angoricity, $\alpha_\tau$ and $\alpha_p$, respectively. 
For example, the shear component (in a two-dimensional system) is  $\Sigma_\tau = (\Sigma_1-\Sigma_2)/2$ and the compressional part is $\Sigma_p = (\Sigma_1+\Sigma_2)/2$, where of  $\Sigma_1$ and $\Sigma_2$ are the eigenvalues of the force-moment tensor $\hat \Sigma_g$. The trace of the force-moment tensor has been commonly referred to as $\displaystyle \Gamma  \equiv \Sigma_p = \mathrm{Tr}~{\hat \Sigma}$. 

\begin{figure}
\includegraphics[width=0.7\linewidth]{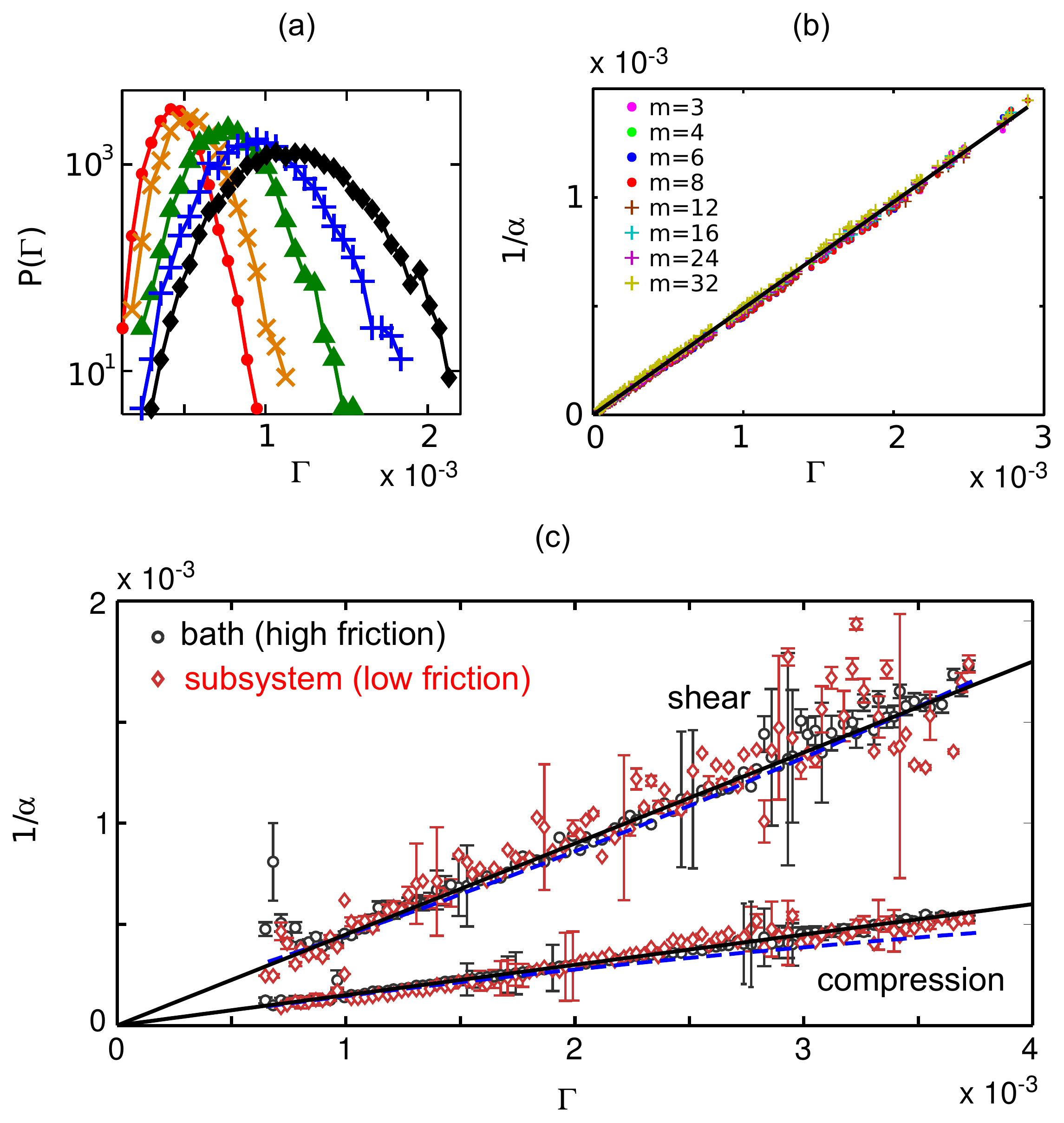}
 \caption{
(a) Histograms of the probability of measuring pressure $\Gamma$, for subsystems of size $m=8$ within simulations of a frictionless, deformable granular material containing 4096 particles~\citep{Henkes2007}. Packing fraction $\phi$ increases from left to right. The ratios of these histograms provide a measurement of the angoricity $\alpha$, using the technique described in Eq.~\ref{e:overlapping2}.
(b) Measured values of the reciprocal-angoricity $1/\alpha$ as a function of $\Gamma$ in the same simulations, showing of slope of 0.49 (solid black line), consistent with theoretical predictions~\citep{Henkes2009}. Results are independent of subsystem size $m$ (colored symbols).
(c) Equilibration of angoricity between a subsystem of low-friction particles (red symbols) within a high-friction bath (black symbols), in repeated realizations of a quasi-two-dimensional packing \citep{Puckett2013}. 
The upper data corresponds to $\alpha_\tau$ (shear), and the lower data $\alpha_p$ (compression). 
The black lines are fits to Eq.~\ref{e:EOS}, with slopes of  0.45 and 0.15, respectively.
Blue dashed lines are for values calculated using the fluctuation dissipation theorem (analogous to Eq.~\ref{e:specificheat}), and are found to be in agreement.
Figures are adapted from \citep{Henkes2007,Puckett2013}. 
\label{f:tests}
}
\end{figure}

The stress ensemble was first tested using numerically-generated  packings of frictionless, deformable particles \citep{Henkes2007}.  As shown in Fig.~\ref{f:tests}a, histograms of $\Gamma$  obtained from clusters of grains within a packing depend strongly on the density of the packing  $\phi$ (proportional to $1/V$ for a fixed number of particles.)  Using the method outlined in Eq.~\ref{e:overlapping1} and \ref{e:overlapping2}, it is possible to (i) test whether the distribution of $\Gamma$ within a packing has a Boltzmann-like form and (ii) extract the differences in angoricity between different packings. This led to the identification of an equation of state: 
\begin{equation}
\frac{1}{\alpha_p} \propto  \langle \Gamma \rangle
\label{e:EOS}
\end{equation}
This equation of state was found to be independent of cluster size (Fig.~\ref{f:tests}b), for clusters more than an order of magnitude smaller than were necessary for compactivity measurements. 
The slope of the observed line is 1/2, in agreement with the value predicted from a configurational entropy estimate of $2/z_\mathrm{iso}$~\citep{Henkes2009}.

A similar technique was applied to packings in experiments \citep{Bi_EPL}, where the presence of friction requires 
considering the full 
tensorial angoricity. A linear equation of state was observed there as well.
These linear equations of state indicate that the configurational entropy of the packings arises via an ``ideal gas'' of non-interacting local stresses. A field theoretical model \citep{Henkes2009} of the spatial stress fluctuations was developed based on numerical data for frictionless disks,  and this has been used to predict the stress fluctuations of numerical and experimental packings under pure shear \citep{Lois2009}.

The generalized Edwards ensemble, combining volume and stress as independent factors, was tested in simulations of three-dimensional (3D) frictionless grains~\citep{Wang2010}. ``Thermalization'' was achieved using independently-generated configurations, and results from the ensemble were found to be in agreement with  predictions from jamming. Experiments in 3D remain hampered by a lack of techniques for measuring vector contact forces in the interior of granular materials. 
No experiments or simulations have yet addressed the Edwards ensemble as a {\itshape joint distribution} of volumes and forces, as has been proposed by \citet{Blumenfeld2012}. 
In addition, little is known about how preparation protocol, particle shape, or particle polydispersity affect the ensembles.

Unlike for compactivity, there has been success in applying angoricity in {\itshape canonical} situations, not just microcanonical ones. 
Fig.~\ref{f:tests}c shows results from the same experiment \citep{Puckett2013} in which compactivity failed to equilibrate. A subsystem of low-friction ($\mu < 0.1$) particles was placed within a bath of higher friction ($\mu = 0.8$) particles. The overlapping histogram method provides angoricities $\alpha_p$ and $\alpha_\tau$ for a set of packings prepared at given volume (and measured pressure $\Gamma$). Each of these two temperature-like quantities equilibrated (independently) between the subsystem and the bath: there is a 0$^\mathrm{th}$ Law.  In addition, each angoricity again had a linear equation of state with the same form as Eq.~\ref{e:EOS}. The coefficients of these equations of state were of similar magnitude to those observed in simulations~\citep{Henkes2007}. However, a configurational entropy calculation \citep{Henkes2009} would predict a slope of $2/z_\mathrm{iso} = 2/3$, which is larger than was observed.

\section{Statistical Mechanics of Jammed States \label{s:statmech}}

As in equilibrium statistical mechanics,  establishing  the generalized Edwards ensemble provides us with a canonical distribution of microstates: $P_{\nu} (X, \hat \alpha)$.  Therefore, the machinery of statistical mechanics can be applied to calculate ensemble averaged quantities.  In particular, one can define the generating function (canonical partition function) :
\begin{equation} 
Z (X, \hat \alpha) = \sideset{}{'}\sum_{ \lbrace {\vec r}_m~, {\vec f}_{mn} \rbrace}  e^{ -\frac{V(\lbrace {\vec r}_m \rbrace)}{X}} e^{ - \hat \alpha: \hat \Sigma(\lbrace {\vec r}_m ~, {\vec f}_{mn}\rbrace)} 
\label{partition}
\end{equation}
where the prime on the summation indicates that the sum over $ \lbrace {\vec r}_m~, {\vec f}_{mn} \rbrace$ is to be restricted to only those states that satisfy the constraints of static mechanical equilibrium for every grain. Imposing these constraints poses difficult challenges
\citep{DeGiuli_thesis, metzger_granular_2004,Metzger2005,Metzger2008}.   While the force and torque balance are equality constraints and are therefore straightforward to implement, the positivity of forces and the static friction law  are both {\itshape inequality } constraints that are notoriously difficult to implement.   

In this section we discuss the different schemes developed to calculate ensemble averaged properties of jammed solids, emphasizing the approximations involved in implementing the constraints of mechanical equilibrium.  In both two (2D) and three dimensions (3D), the constraints of force balance can be enforced through the introduction of gauge fields \citep{DeGiuli2011,DeGiuli_thesis,Henkes2009}.  In 2D, these gauge fields have an elegant geometric representation that has been used in recent years to calculate statistical properties of jammed states at various levels of approximations.    This section is organized as follows:  we first focus on how the intertwining of the volume and stress in the partition function. Eq. \ref{partition} has been treated in the literature. Then, we discuss methods applicable in 2D before providing a brief description of a similar mathematical formulation in 3D, along with the difficulties that arise when moving from 2D to 3D. 

Eq.  \ref{partition} is a function of both $X$ and $\hat \alpha$, and this cannot, in general, be written as a product: $Z(X, \hat \alpha) \neq Z(X) Z(\hat \alpha)$.  Recent work~\citep{Blumenfeld2012}  draws attention to this fact and obtains an expression for the partition function by combining the loop force formalism~\citep{Ball2002} (to enforce force and torque balance in 2D) with the quadron method for taking into account the volume degrees of freedom. 
For the special case of an isostatic assembly of grains, where each geometry leads to a unique configuration of contact forces, the calculations \citep{Blumenfeld2012}  give rise to an equipartition-like equation of state linking mean volume $\langle V\rangle$ and compactivity  $\langle V \rangle = \left(zN/2 + M\right)X$, where the $zN/2$ structural degrees of freedom and the $M$ independent boundary force degrees of freedom contribute equally.

Most statistical treatments of jammed states have relied on some  form of decoupling of the volume and stress terms in the partition function.  The original Edwards ensemble is an extreme example: the stress fluctuations were not considered at all. 
Conversely, the force network ensemble (FNE)~\citep{snoeijer03} is based on a complete decoupling of the geometric and force degrees of freedom.  It is a microcanonical ensemble that takes a flat measure on all allowed force states for a given, {\it fixed}  geometric network.  The FNE decoupling is based on a separation of scales: for infinitely rigid grains or packings of soft grains very close to jamming, small geometric displacements lead to large changes in the interparticle contact forces. The FNE is particularly suited to computing probability distributions $P(f)$ of single contact forces $f$ in both lattice geometries and some off-lattice geometries~\citep{Snoeijer2004Ensemble-theory,tighe05,van_eerd_tail_2007}.    It has also proved useful  in understanding force fluctuations in frictional packings~\citep{unger_force_2005}.  There are many excellent reviews of the application of the FNE and we refer the reader to these for a comprehensive discussion \citep{Tighe:2010ff,van_eerd_numerical_2009}.

Another variation on the decoupling theme is to consider a canonical  stress ensemble which neglects all volume fluctuations~\citep{Chakraborty2010,Henkes2007,Bi_EPL}. This is the analog of the $(T,V,N)$ ensemble of equilibrium statistical mechanics.  This is distinct from the FNE since both positions and forces are retained as degrees of freedom through the force moment tensor.   The stress ensemble is most appropriate for describing grains subject to shear stresses at constant volume~\citep{Bi_EPL}.   It is a reasonable approximation in situations where volume fluctuations are negligible compared to stress fluctuations.  The experimental tests involving both $X$ and $\hat \alpha$, described in the previous section, suggest that $\hat \alpha$ can be defined even in cases where the definition of $X$  is murky.  In the remainder of this section, therefore,  we focus on the pure stress ensemble (which includes the FNE as a special case).  The interdependence of volume and stress fluctuations in the 
generalized ensemble, which is the analog of the $(T,P,N)$ ensemble, has received less attention~\citep{Blumenfeld2012}.   One of the important questions in this regard is the range of 
volumes over which jammed packings can be created \citep{ciamarra_statistical_2012}.

\subsection{2D Granular Solids}
\begin{figure}[htbp]
\begin{center}
\includegraphics[width=1\columnwidth]{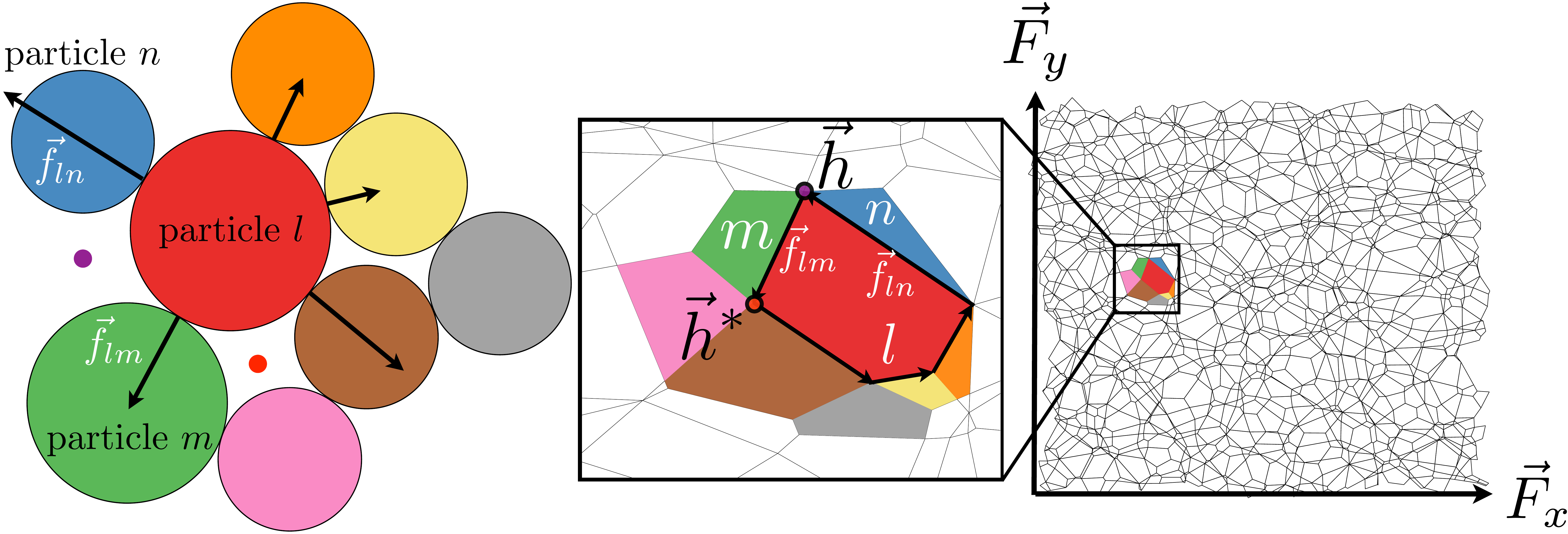}
\caption
{Force tilings for a frictionless grain packing from simulations: The first  panel shows grain $l$  supported by balanced contact forces exchanged with its neighbors ($m$, $n$ etc.).  The real space contact network defines voids such as the one bordered by $m$, $l$, $n$, (purple dot) on which the vector loop forces, ${\vec h}$, ${\vec h^*}$ live.  Going around a grain in a counterclockwise direction, the height is incremented by the contact force separating two voids.   The second panel shows the force tile corresponding to grain $l$ and its neighbors. The \emph{force vectors} define the edges of the \emph{force tiles}, the faces are labeled by grain indices, and  the heights ${\vec h}$ and ${\vec h^*}$ become the vertices of the tiling. Because of action-reaction and force balance, force tiles tesselate height space as shown in the third panel. The boundaries  of the tesselation are defined by the vectors ${\vec F}_x$ and ${\vec F}_y$ 
that define the force moment 
tensor $\hat \Sigma$ (Eq.
 \ref{f_sigma_relation}). }
\label{FT_example}
\end{center}
\end{figure}

In 2D, the constraints of local force balance (Eq.~\ref{divergencefree}) allows for the definition of a vector gauge field, ${\vec h}(x,y)$ \citep{Henkes2005,Henkes2009,DeGiuli_thesis,DeGiuli2011} such that\footnote{\label{note1} The generalized definition of the curl used implies that $ \sigma_{\mu \nu} =\sum_{\mu' \nu' } \epsilon_{\mu \mu'  \nu' }  \partial_\mu'  W_{\nu'  \nu}$ for 3D and $\sigma_{\mu \nu} = \sum_{\mu '} \epsilon_{\mu \mu'} \partial_{\mu'} h_{\nu}$ in 2D, where $\epsilon$ is the Levi-Civita symbol}
\begin{equation}
\hat{\sigma}=  \vec{\nabla} \times \vec{h}.
\label{height_defn}
\end{equation}
In a discrete formulation at the grain level, this single-valued, vector height field is defined on the dual space of voids in a granular packing\ citep{Ball2002}, as illustrated in  Fig. \ref{FT_example}. Going around a grain in a counterclockwise direction, the height field gets incremented by the contact force separating two voids.  The mapping to this dual space of loops and heights is rigorous and retains all of the information about the original packing \citep{Henkes2009,DeGiuli_thesis,DeGiuli2011,Ball2002}.

 A simpler geometric representation that does not carry information about 
the real-space geometry (but retains the topology of the contact network and accurately represents the structure in height-space) is the tiling picture~\citep{Tighe:2008ye}  shown in the left panel of  Fig.~\ref{FT_example}.  Since the forces on every grain have to add up to zero, each grain can be represented by a polygonal tile constructed by connecting  the force vectors head-to-tail.  Newton's $3^\mathrm{rd}$ Law requires that the edges of tiles exactly coincide, because each edge (contact force) is shared by two grains.  The vertices of this tiling are the heights, and the faces of the tiles represent grains.    
Since  the force associated with touching grains are equal in magnitude and opposite in direction,  force tiles tesselate height space and  the vertices of all force tiles fill a parallelogram (see Fig.~\ref{FT_example}) in height space spanned by two vectors $\vec{F}_x$ and $\vec{F}_y$ defined in Eq. \ref{f_sigma_relation}.

The force tilings can form the basis for calculating partition functions and ensemble averages.  In the tiling representation,  the microcanonical stress ensemble is comprised of all tilings by $N_f$ tiles of a parallelogram bounded by ${\vec F}_x$ and ${\vec F}_y$, with each tile corresponding to a force-bearing grain with more than two contacts.  Grains with two contact forces or fewer are not represented in a tiling.   We will discuss this aspect in the context of tiling corresponding to frictional grain packings.  A canonical stress ensemble would be defined by $\hat \alpha$ and in this ensemble the boundary vectors defining the parallelogram would be allowed to fluctuate.  In order to perform an actual calculation, the microscopic probabilities, $\omega_{\nu}$ of each tiling would have to be defined and the additional constraints of torque balance and  static friction would have to be imposed on the tilings.   The latter two are irrelevant for frictionless grains, and all calculations (to date)  of 
statistical properties of frictionless grains  have assumed equiprobability $\omega_{\nu}=1$.   For frictionless grains, the FNE has been used successfully to calculate statistical properties of grains in lattice geometries \citep{tighe_stress_2011}.   Interestingly, tilings of frictionless grains exhibit an additional extensive quantity that modifies the stress ensemble.  In the rest of this section, we summarize the application of stress ensembles to assemblies of frictionless and frictional  grains. 

\subsubsection{Frictionless Grains} For dry granular solids, the forces between grains are purely repulsive. Without friction and for disk-shaped particles, these contact forces are all central forces.  It can be easily proven that if forces are frictionless and repulsive, all force tiles must strictly be convex in shape.  When all force tiles are convex in shape, they form a planar graph in height space. As illustrated in Fig.~\ref{FT_example}, the result is equivalent to the Maxwell-Cremona reciprocal tiling or force tiling \citep{Tighe:2008ye}, where each grain is represented by a polygonal tile. Hence, the area of the parallelogram spanned by $\vec{F}_x$ and $\vec{F}_y$ is simply the total area taken up by all tiles in height space
\begin{equation}
A_\mathrm{tot} = \sum_{m} a_m = \left\vert \vec{F}_x \times \vec{F}_y \right\vert ~,
\label{area}
\end{equation}
where, the last equality holds only in the thermodynamic limit or for periodic boundary conditions~\citep{tighe_stress_2011}. Eq. \ref{area} demonstrates that there is an additional extensive quantity for 2D, frictionless assemblies of grains  that is distinct from $\hat \Sigma = \sum_m \hat{\sigma}_m$. While the conservation principle is a prerequisite for the additivity and conservation of $A_{tot}$, in general, it does not imply so.
In the same way as Eq.~\ref{partition} is derived, a Boltzmann term involving the force tile area must also be added to as a result of the area constraint to the microstate probability for frictionless grains: 
 \begin{equation}
P_{\nu} = \frac{1}{Z} \exp \left(-\hat{\alpha}:\hat{\Sigma}_{\nu}- b \mathcal{A}_\nu\right ),
\label{full_se}
\end{equation}
where $\mathcal{A}_\nu$ is the area of a microscopic state and 
\begin{equation}
b = \frac{\partial S}{\partial A_{tot}}.
\end{equation}
For isotropic states,  $\hat{\alpha}:\hat{\Sigma}_{\nu}$ reduces to $\alpha_p \Gamma_\nu$, where $\Gamma_\nu$ is the trace of the force moment tensor. $\Gamma_\nu$ is  proportional to the perimeter of a tile, $\mathcal{P}$. Furthermore, for a polygon that is convex and isotropic in shape, area is related to perimeter via $\mathcal{A} \propto \mathcal{P}^2$. 
(Conversely, anisotropic polygons or complex polygons may have power $<2$.)
In this case, Eq.~\eqref{full_se} becomes 
 \begin{equation}
P(\mathcal{P}) \propto \frac{1}{Z} \exp(- \alpha_p \mathcal{P} -  \beta c {\mathcal P}^2 ),
\end{equation}
which coincides with the result obtained in a test using FNE \citep{Tighe:2008ye} and entropy maximization.
Within the tiling framework, force and pressure distributions can also be calculated using the FNE on a lattice and off-lattice with excellent agreement between theory and Monte Carlo simulations~\citep{Tighe:2008ye,tighe_stress_2011,tighe_force_2010-1}, though the approach remains to be tested on naturally generated packings with coupled forces and geometry.  One particularly interesting result is that the distribution of the magnitude of individual contact forces, $P(f)$ is Gaussian for packings of frictionless grains~\citep{Tighe:2008ye,tighe_stress_2011,tighe_force_2010-1}, and that this is direct consequence of the additional area conservation.   The absence of this additional conservation principle in frictional grains, to be discussed below,  raises the intriguing possibility that the exponential form of $P(f)$ observed in experiments is connected to force chains 
and friction.

\subsubsection{Frictional Grains \label{s:friction_stat_mech}}  
\begin{figure}[htbp]
\begin{center}
\includegraphics[width=0.8\columnwidth]{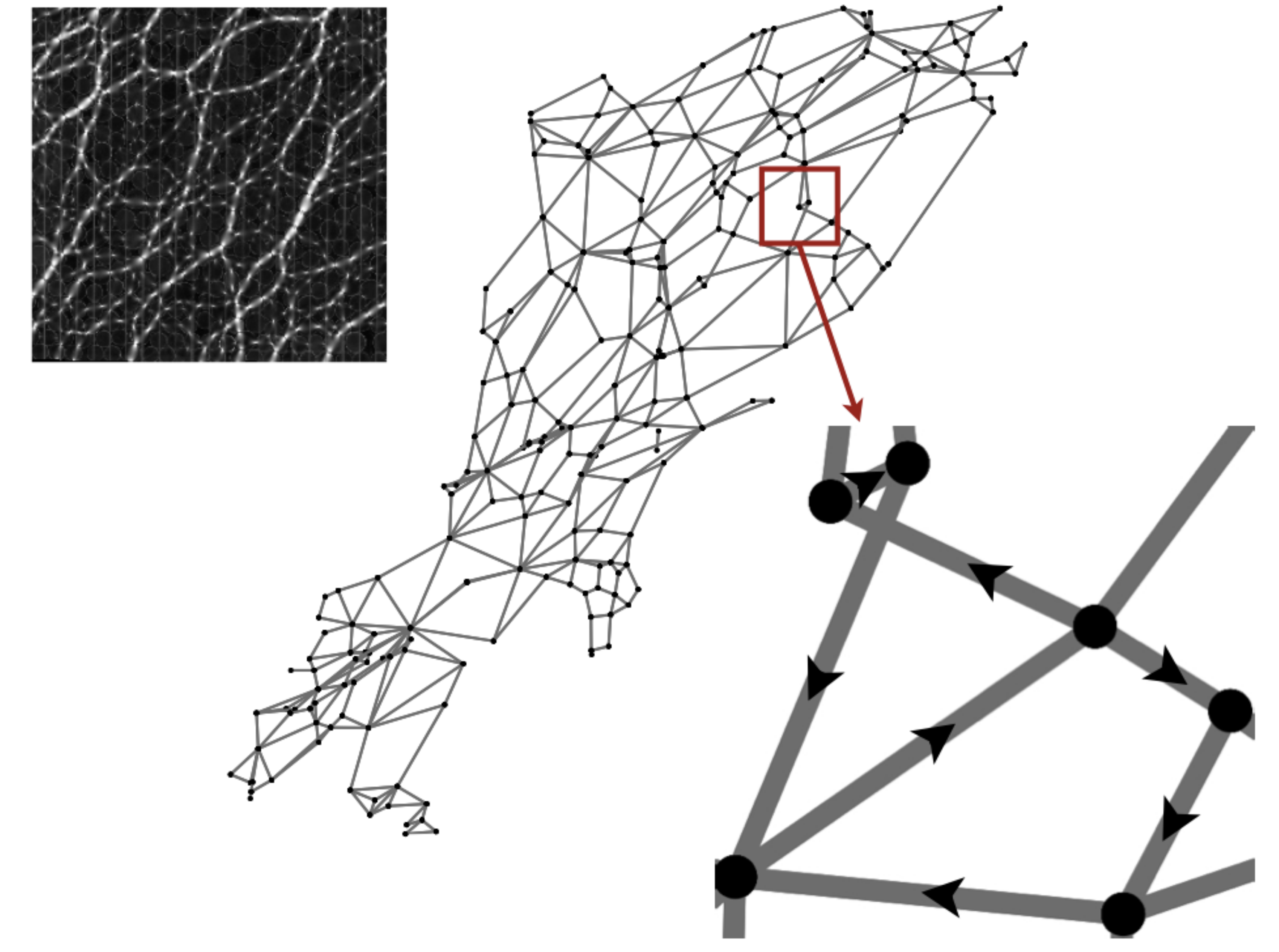}
\caption{Force tilings constructed from experimental shear-jammed states.  Clockwise from left:  A photo elastic image \citep{Jie_Ren_Thesis} showing force chains, force tiling corresponding to this image, and a part of the force tiling showing a self-intersecting polygon: the arrows mark the direction of contact forces}
\label{tiling_frictional}
\end{center}
\end{figure}

The tiling framework for frictional grains admits non-convex and even complex (self-intersecting) polygons.   Fig. \ref{tiling_frictional} shows a tiling constructed from experimental data on jammed states created through shear \citep{Sarkar2013}.  Grains located along force chains, where there is a well defined direction of large forces, tend to give rise to self-intersecting polygons, as illustrated in Fig. \ref{tiling_frictional}.  The additional area conservation is therefore not applicable to tilings of frictional grains, which are non-planar because of the presence of self-intersecting polygons.  For frictional grains, therefore, the generalized Edwards ensemble encodes all the conservation laws, and there are no additional conserved quantities.  The number $N_f$ of tiles corresponds to grains with more than two contacts.  In frictionless grains, this number is not far from the actual number of grains.  However, in frictional systems at 
low 
densities 
close to random loose 
packing, force-bearing grains organize into force-chains that meander through a host of spectator particles that bear essentially no force \citep{Bi2011ShearJam,Cates1998}.  For these systems, $N_f$ can be very different from the actual number of grains.   What the tiling framework shows is that for mechanical stability, what matters is $N_f$.  The volume in the original Edwards ensemble refers to the volume occupied by {\it all} grains.  

 In order to perform calculations for frictional grains, assuming equiprobability of microstates, the constraints of torque balance and static friction would have to be imposed explicitly on the sampling of tiles.   To our knowledge, such a calculation has not been attempted yet.  A meanfield calculation has been performed \citep{DeGiuli_thesis} and leads to an approximation of the (force space) configurational entropy and hence a prediction for the equation of state for grains with friction.   In another development, force tilings constructed from experimental data have been analyzed to demonstrate that emergence of rigidity in shear-jammed solids~\citep{Bi2011ShearJam} can be related to a persistent pattern in the density of vertices in force tiles \citep{Sarkar2013}.    
 
There is a clear need for the development of a Monte Carlo, Metropolis algorithm that samples tiling according to the Boltzmann-like distribution of the stress ensemble, but which also imposes the constraints of torque 
balance and the Coulomb friction law $|f_t| \le \mu |f_N|$ on the sampling.   The advantage 
that the 
tiling framework offers is the ability to encode these constraints in the geometrical language of polygons.   It should be emphasized that sampling tilings is not restricted to a fixed geometry of grains in real space.  What is fixed is the number of force-bearing grains, their positions and contact forces are dynamical variables in a Monte Carlo scheme, for example.  The positions do not enter the tiling picture explicitly, and an assumption being made in sampling tilings without connecting to the real-space network is that there is real-space network of grains that corresponds to every tiling.   This assumption is reasonable if there is a decoupling of forces and positions, and the indeterminacy arising from tangential forces helps in this regard.     Even if numerical calculations are difficult, the tiling framework allows us to construct correlations in height space, which are often more illuminating than real-space correlations~\citep{Sarkar2013}. This is because the primary source of correlations in 
dry 
granular ensembles are the constraints of mechanical equilibrium, and these are explicitly taken into account in all tiling configurations.

\subsection{Three dimensions \label{3D}}

In the three-dimensional case, the definition of an equivalent microscopic framework to the two-dimensional tiling becomes much more complex~\citep{DeGiuli_thesis,DeGiuli2011}. At the continuum level, the equivalent to the vector gauge field $\vec{h}$ is a tensor field $\hat{W}$ such that Eq.~\ref{height_defn} becomes\textsuperscript{\ref{note1}} $\hat{\sigma}=\vec{\nabla} \times \hat{W}$. Equivalently to the two-dimensional case, we can use Stokes' theorem to show that the stress in a volume is a boundary term; the surface integral of $\hat{W}$ dotted with the unit normal.
Since $\hat{W}$ is a tensor, rather than a vector, it is unclear what the analogue of the dual tiling space would be; no 3D equivalent of the tiling has been proposed so far.
Recently, \citet{DeGiuli2011} derived a microscopic interpretation of $\hat{W}$, analogous to the two dimensional height field. However, while the microscopic variables are vectorial, they only map to $\hat{W}$ by incorporating significant parts of the local geometry. 
Finally, it should be noted that for realistic three dimensional packings gravitation needs to be incorporated, and the framework based on conservation of $\hat \Sigma$  needs to be extended to include gravity as an external field.


\section{Application to Dynamics \label{s:dynamics}} 

As mentioned in the introduction, one of Edwards' original ideas was that the dynamics of slowly-driven granular materials can be understood based on the microstate probabilities of jammed states.  This seems plausible if the driving is slow enough that there are rare transitions from one jammed state to another, and the transition probabilities are controlled by properties of the jammed states.  This picture is reminiscent of  frameworks constructed for describing the glassy dynamics of thermal systems such as the trap model \citep{Bouchaud-trap}, the soft-glassy-dynamics (SGR) models \citep{Sollich1998} and the shear-transformation-zone (STZ) theory \citep{Langer2008Shear-transform}.
In all of these models, there are transitions into and out of trap-like entities such as free-energy minima.  
For example, in the SGR model it is assumed that the material can be partitioned into mesoscopic regions. Each region is able to  accommodate some elastic energy and strain. Plastic deformations of the mesoscopic region occur either due to (i) global driving or
(ii) fluctuations due to the elastic energies released by other mesoscopic regions undergoing plastic deformation \citep{Sollich1998}.  In this meanfield model, the latter  is represented as an activated process with an effective temperature that is different from the bath temperature \citep{Sollich1998}.

A natural avenue of exploration of  granular dynamics based on the ensemble ideas is to adopt the same philosophy as SGR, and assume that angoricity plays the role of the effective temperature.   The idea that self-activated processes induced by stress fluctuations, or mechanical noise,  influence the rheology of slow granular flows has been explored in the literature \citep{Reddy:2011qf,Kamrin}, and it seems reasonable to assume that  angoricity  determines the strength of this  mechanical noise.  A angoricity-based, SGR-like framework has been used to analyze stress fluctuations in laboratory granular Couette flows \citep{Bi2008,Bi2009}. 
This formalism provides an explanation for the observed logarithmic strengthening \citep{HartleyNature} of granular materials.  A key point is that one needs to use a stress ensemble rather than a energy ensemble (as would be the case for SGR), in order to obtain the correct scaling with shear rate.  Such applications of the  stress ensemble to analyze granular rheology are in their infancy, and much remains to be explored.


\section{Summary and Open Questions \label{s:questions}}

Our overview of the application of statistical ensemble ideas to athermal, granular materials highlights recent advances in this area.    One of our primary objectives has been to emphasize the similarities and differences between athermal and thermal ensembles.   The experimental and numerical tests discussed in this review illustrate the validity of the ensemble framework  in several different contexts.   The equilibration of angoricity, which is the analog of thermalization in the stress ensemble, has been shown to work both in experiments and simulations.   Measurements of angoricity and compactivity based on the method of overlapping histograms  (Eq. \ref{e:overlapping2}), and on fluctuation-response relations such as Eq. \ref{e:specificheat}, seem to agree with each other, which suggests that even in these athermal ensembles, fluctuations within a system determine its linear response to driving.

The majority of tests have been performed in 2D, and the applicability of ensembles to 3D packings is much less clear.   Equations of state such as those relating the angoricity to the external stress and compactivity to volume have been measured in only a few systems.  In particular, equations of state involving the full tensorial angoricity have remained largely unexplored, as have equations of state relating compactivity and angoricity.   Since equiprobability does not seem to be universally valid for jammed systems, equations of state will depend upon the probabilities $\omega_{\nu}$ of the microstates, and these are expected to be sensitive to preparation protocols, friction coefficients and particle shapes.   The variation in equations of states is, therefore, expected to be greater than that found in thermal ensembles.  Validity of the ensemble framework, however, narrows down the choice of macrosopic variables that should enter an equation of state.   

From the perspective of calculations based on ensembles, the difficulty of imposing the constraints of mechanical equilibrium has hindered progress.   In particular, calculations of partition functions have been hampered by two inequality constraints: positivity of forces and the Coulomb condition for static friction.    The situation is hopeful in 2D, where the constraints have elegant geometrical representations, but little is know about how to proceed in 3D.

Within the context of dynamics, the ensemble framework provides a natural basis for stress-induced fluctuations being represented by a type of thermal noise.  The rheology of slowly driven systems could potentially be completely linked to angoricity and compactivity by generalizing the idea of thermally activated processes to stress-activated processes that lead to plastic failure.   If dynamics can be related to ``granular equilibrium'' concepts such as angoricity and compactivity, that would be a big step since granular rheology could then be related to stresses imposed at the boundaries.

To summarize,  the last decade has seen much progress in testing and establishing the generalized Edwards ensemble, which establishes the laws of granular equilibrium.  We should emphasize that the ensemble framework is relevant only to static and slowly-driven granular materials where the kinetic energy of the grains do not play a significant role.    For systems that are governed by the laws of  granular equilibrium, stress has emerged as the physical property that dominates collective organization.


\section{Acknowledgements\label{s:ack}}

The force tilings in Figs.~\ref{FT_example} and \ref{tiling_frictional} were created by Sumantra Sarkar.
We are grateful for support from the National Science Foundation under grants DMR-0549762 and DMR-0905880 (BC, DB, and SH), the Brandeis IGERT (DB),  and DMR-0644743 and DMR-1206808 (KD).
The authors acknowledge useful discussions with Bob Behringer, Olivier Dauchot, Daan Frenkel, Mitch Mailman, Corey O'Hern, James Puckett, Sriram Ramaswamy, Matthias Schr\"oter, and Brian Tighe.


\clearpage


\end{document}